\documentclass[aps,prb,preprint]{revtex4}
\usepackage{graphicx}
\usepackage{dcolumn}
\usepackage{amsmath}
\usepackage{amssymb}
\def\uu{\uparrow\uparrow}
\def\ud{\uparrow\downarrow}
\def\rv{{\bf r}}

\def\Rv{{\bf R}}

\def\beq{\begin{equation}}
\def\eeq{\end{equation}}

\def\erf{{\rm erf}}
\def\refcite{\onlinecite}

\begin{document}

\title{Kohn-Sham calculations combined with an average
pair-density functional theory}

\author{Paola Gori-Giorgi$^*$ and Andreas Savin}

\affiliation{Laboratoire de Chimie Th\'eorique, CNRS
UMR7616, Universit\'e Pierre et Marie Curie, 4 Place Jussieu,
F-75252 Paris, France \\
$^*$E-mail: paola.gori-giorgi@lct.jussieu.fr\\
www.lct.jussieu.fr/pagesequipe/DFT/gori}

\begin{abstract}
A recently developed 
formalism in which Kohn-Sham calculations are combined
with an ``average pair density functional theory'' is 
reviewed, and some
new properties of the effective electron-electron interaction
entering in this formalism are derived. A preliminary construction
of a fully self-consitent scheme is also presented in this framework.
\end{abstract}
\maketitle

\section{Introduction}
Kohn-Sham (KS) Density Functional Theory\cite{kohnnobel,science,FNM} 
(DFT) is nowadays one of the most
popular methods for electronic structure calculations both in chemistry
and solid-state physics, thanks to its combination
of low computational cost and reasonable performances. 
The accuracy of a KS-DFT result is limited
by the approximate nature of the exchange-correlation energy
density functional $E_{xc}[n]$. Typical cases in which 
present-day DFT fails are strongly correlated systems,
the description of van der Waals forces,  the handling of near degeneracy.
Much effort is put nowadays in trying to improve the DFT performances via
the construction of better approximations 
for the KS $E_{xc}[n]$ (for recent reviews, see, e.g., 
Refs.~\refcite{science,FNM,prescription}), 
or via alternative routes, like, e.g., the use of
non-KS options.\cite{varieLRSRDFT} A popular
trend in the development of new KS $E_{xc}[n]$
is the use of 
the exact exchange functional $E_x[n]$ (in terms of the KS orbitals), and
thus the
search for an approximate, compatible, correlation functional $E_c[n]$.

In this work we review the basis of a theoretical framework~\cite{GS1,GS2}
 in which
KS-DFT is combined with an ``average pair density functional theory'' (APDFT)
that provides an explicit construction for $E_{c}[n]$, transfering
the work of finding an approximate functional to the
search of an effective particle-particle interaction.
A self-consitent scheme for this approach is presented, 
and some new properties of the effective
interaction that enters in this combined formalism are derived.
Very preliminary applications are discussed.

\section{Definitions}
\label{sec_def}
Our target problem is finding the ground-state energy of 
the standard $N$-electron hamiltonian in the
Born-Oppenheimer approximation (in Hartree atomic units,
$\hbar=m=a_0=e=1$, used throughout),
\begin{eqnarray}
H & =& T+V_{ee}+V_{ne}, \label{eq_Hphys} \\
T & = & -\frac{1}{2}\sum_{i=1}^N\nabla_i^2, \\
V_{ee} & = & \frac{1}{2}\sum_{i\neq j}^N \frac{1}{|\rv_i-\rv_j|}, \\
V_{ne} & = & \sum_{i=1}^N v_{ne}(\rv_i),
\end{eqnarray}
where $v_{ne}$ is the external potential due to nuclei.
Given $\Psi$, the exact ground-state wavefunction of $H$, we consider
two reduced quantities that fully determine, respectively,
 the expectation values
$\langle\Psi|V_{ne}|\Psi\rangle$ and $\langle\Psi|V_{ee}|\Psi\rangle$, i.e.,
the electronic density,
\beq
n(\rv)=N\sum_{\sigma_1...\sigma_N}\int |\Psi(\rv\sigma_1,\rv_2\sigma_2,...,
\rv_N\sigma_N)|^2d\rv_2...d\rv_N,
\eeq
and the spherically and system-averaged pair density $f(r_{12})$ (APD), which is
obtained by first considering the pair density $P_2(\rv_1,\rv_2)$,
\beq
P_2(\rv_1,\rv_2)=N(N-1)\sum_{\sigma_1...\sigma_N}
\int |\Psi(\rv_1\sigma_1,\rv_2\sigma_2,\rv_3\sigma_3,...,\rv_N\sigma_N)|^2 d\rv_3...d\rv_N,
\label{eq_defP2}
\eeq
and then by integrating it over all variables except
$r_{12}=|\rv_1-\rv_2|$,
\beq
f(r_{12}) = \frac{1}{2}
\int P_2(\rv_1,\rv_2)\frac{d\Omega_{\rv_{12}}}{4\pi} d\Rv,
\label{eq_deff}
\eeq
where
$\Rv=\frac{1}{2}(\rv_1+\rv_2)$, $\rv_{12}=\rv_2-\rv_1$.
The function $f(r_{12})$ is also known in chemistry as intracule
density~\cite{thakkar,Coulson,Cioslowski1,Cioslowski2,ugalde1,davidson,coleman}, and,
in the special case of a an electron liquid of uniform density $n$, 
is related to the radial pair-distribution function $g(r_{12})$
 by $g(r)=2 f(r)/(n N)$. We thus have
\begin{eqnarray}
\langle\Psi|V_{ne}|\Psi\rangle  =  \int n(\rv) v_{ne}(\rv) d\rv 
\label{eq_vnefromn}\\
\langle\Psi|V_{ee}|\Psi\rangle  =  \int \frac{f(r_{12})}{r_{12}} d\rv_{12}=
\int_0^{\infty}\frac{f(r_{12})}{r_{12}}4\pi r_{12}^2 dr_{12}.
\label{eq_veefromf}
\end{eqnarray}

In the following text we will also deal with modified systems in which 
the external potential
and/or the electron-electron interaction is changed. Thus, 
the notation $V_{ne}$ and $V_{ee}$ will
be used to characterize the physical system, while the modified systems 
will be defined by  
$V=\sum_{i=1}^N v(\rv_i)$ and $W=\frac{1}{2}\sum_{i\neq j}^N w(|\rv_i-\rv_j|)$, 
where the pairwise interaction $w$ always depends only on
$|\rv_i-\rv_j|$.

\section{The exchange-correlation functional of KS-DFT}
\label{sec_DFT}
In standard DFT one defines a universal functional of the one-electron density
$n(\rv)$ as resulting from a constrained search over all the antisymmetric
wavefunctions $\Psi$ that yield $n$~\cite{levy}
\beq
\tilde{F}[n;V_{ee}]=\min_{\Psi \to n} \langle \Psi |T+V_{ee}|\Psi\rangle,
\label{eq_dftlevy}
\eeq
or, more completely, as a Legendre transform~\cite{lieb}
\beq
F[n;V_{ee}] = \sup_v\left\{\min_{\Psi} \langle \Psi |T+V_{ee}+V|\Psi\rangle
 -\int n(\rv) v(\rv) d\rv\right\}.
\label{eq_dftlieb}
\eeq
In both Eqs.~(\ref{eq_dftlevy}) and~(\ref{eq_dftlieb}), the dependence
on the electron-electron interaction
 has been explictly shown in the functional. The universality of the functional
$F$ stems exactly from the fact that the e-e interaction is always
$1/r_{12}$.
The ground-state energy $E_0$ of the system can then be obtained by minimizing 
the energy with respect to $n$,
\beq
E_0=\min_n\left\{F[n;V_{ee}]+\int n(\rv) v_{ne}(\rv) d\rv\right\}.
\label{eq_Edft}
\eeq

A possible way to derive the Kohn-Sham equations in DFT is to   
define a set of hamiltonians
depending on a real parameter $\lambda$~\cite{gunnarsson,wang,adiabatic},
\beq
H^\lambda=T+W^\lambda+V^\lambda,
\label{eq_Hlambda}
\eeq
having all the same one-electron density, equal to the one
of the physical system 
\beq
n^\lambda(\rv)=n(\rv)\qquad \forall \lambda.
\label{eq_nconst}
\eeq
If $W^{\lambda_{\rm phys}}=V_{ee}$ and $W^{\lambda=0}=0$
(e.g., $W^\lambda=\lambda V_{ee}$), one 
can slowly switch off the electron-electron interaction, while keeping the
density fixed via a suitable external potential $V^\lambda$.
Obviously, the APD $f(r_{12})$ changes with $\lambda$.
By the Hellmann-Feynmann theorem,
\beq
\frac{\partial E_0^{\lambda}}{\partial \lambda}   =  
\langle \Psi^{\lambda} |\frac{\partial W^{\lambda}}{\partial \lambda}+
\frac{\partial V^{\lambda}}{\partial \lambda}|\Psi^{\lambda}\rangle 
= \int f^{\lambda}(r_{12})\frac{\partial w^{\lambda}(r_{12})}
{\partial \lambda}d\rv_{12}+
 \int n(\rv)\frac{\partial v^{\lambda}(\rv)}
{\partial \lambda}d\rv,
\label{eq_HFDFT}
\eeq
so that by directly integrating Eq.~(\ref{eq_HFDFT}), and by combining
it with Eq.~(\ref{eq_Edft}), one obtains
\beq
F[n;V_{ee}]=T_s[n]+\int_0^{\lambda_{\rm phys}} d\lambda\int d\rv_{12}
f^{\lambda}(r_{12})\frac{\partial w^{\lambda}(r_{12})}
{\partial \lambda},
\label{eq_defTs}
\eeq
where $T_s[n]=F[n;0]$ is the kinetic energy of a noninteracting system
of $N$ spin-$\frac{1}{2}$ fermions with density $n(\rv)$. The adiabatic
connection in DFT thus naturally defines the Kohn-Sham non-interacting kinetic
energy functional $T_s[n]$. The second term in the right-hand-side
of Eq.~(\ref{eq_defTs}) is an exact expression, in terms of the APD
$f^\lambda(r_{12})$, for the Hartree and the exchange-correlation
functional, $E_H[n]+E_{xc}[n]$. The one-body potential at $\lambda=0$
 is the familiar Kohn-Sham potential, $v^{\lambda=0}(\rv)=v_{\rm KS}(\rv)$.

The traditional approach of DFT to construct approximations for $E_{xc}[n]$
is based on the idea of universality. For example, the familiar local-density
approximation (LDA) consists in transfering,
in each point of space,
the pair density from the uniform electron gas to obtain
an approximation for $f^\lambda(r_{12})$ in Eq.~(\ref{eq_defTs}). 
Our aim is to develop an alternative strategy in which realistic
APD $f^\lambda(r_{12})$ along the DFT adiabatic connection are constructed
via a formally exact theory that must be combined with the KS equations
in a self-consitent way. The formal justification for this
``average pair-density functional theory'' (APDFT) is the object of
 the next Sec.~\ref{sec_APDFT}. 

\section{Average pair density functional theory}
\label{sec_APDFT}
As shown by Eqs.~(\ref{eq_vnefromn}) and (\ref{eq_veefromf}), the APD
$f(r_{12})$ couples to the operator $V_{ee}$ in the same way as the
electronic density $n(\rv)$ couples to $V_{ne}$. 
In order to derive an ``average pair density functional theory'' (APDFT)
we thus simply repeat the steps of the previous Sec.~\ref{sec_DFT}
by switching the roles of $f$ and $n$, and of $V_{ee}$ and 
$V_{ne}$.\cite{GS2}

We thus define a system-dependent functional (i.e., a functional
depending on the external potential $V_{ne}$, and thus on the specific system) 
of the APD $f(r_{12})$ as
\beq
\tilde{G}[f;V_{ne}]=\min_{\Psi \to f} \langle \Psi |T+V_{ne}|\Psi\rangle,
\label{eq_fftlevy}
\eeq
where, again the minimum is over all antisymmetric wavefunction $\Psi$ that yield
a given $f(r_{12})$. We can also define the system-dependent functional $G$ as
\beq
G[f;V_{ne}] =  \sup_w\left\{\min_{\Psi} \langle \Psi |T+W+V_{ne}|
\Psi\rangle  -\int f(r_{12}) w(r_{12}) d\rv_{12}\right\}.
\label{eq_fftlieb}
\eeq
The ground-state energy could then be obtained as
\beq
E_0=\min_{f\in {\mathcal N}_f}\left\{G[f;V_{ne}]+\int \frac{f(r_{12})}{r_{12}} d\rv_{12}\right\},
\label{eq_Efromf}
\eeq
where ${\mathcal N}_f$ is the space of all $N$-representable APD (i.e., coming
from the contraction of an $N$-particle antisymmetric wavefunction). The
definition of the space ${\mathcal N}_f$ is evidently related to the 
$N$-representability conditions for the pair density, for which recent 
interesting progresses have been made~\cite{ayersNPD}. In our case,
however, we combine APDFT with DFT so that the minimization of
Eq.~(\ref{eq_Efromf}) is never directly carried on.

In order to find the analog of the KS system for APDFT, we 
define an adiabatic connection similar to the one of Eq.~(\ref{eq_Hlambda})
in which, this time, we switch off the external potential. We
thus introduce a set of hamiltonians depending on a real parameter 
$\xi$,
\beq
H^\xi=T+W^\xi+V^\xi,
\label{eq_Hxi}
\eeq
in which the function
$f(r_{12})$ is kept fixed, equal to the one of the physical
system,
\beq
f^\xi(r_{12})=f(r_{12})\qquad \forall \xi.
\label{eq_fconst}
\eeq
If $V^{\xi_{\rm phys}}=V_{ne}$ and $V^{\xi=0}=0$
(e.g., $V^\xi=\xi V_{ne}$), we are
switching continuously from the physical system, 
to a system of $N$ free electrons
interacting with a modified potential $w^{\xi=0}(r_{12})$. 
That is, $f(r_{12})$ is kept
fixed as $\xi$ varies by means of a suitable electron-electron
interaction $W^\xi$ while the one-electron density $n(\rv)$
changes with $\xi$.
Again, by the Hellmann-Feynmann theorem, we find
\beq
\frac{\partial E_0^{\xi}}{\partial \xi}   =  
\langle \Psi^{\xi} |\frac{\partial W^{\xi}}{\partial \xi}+
\frac{\partial V^{\xi}}{\partial \xi}|\Psi^{\xi}\rangle =
\int f(r_{12})\frac{\partial w^{\xi}(r_{12})}
{\partial \xi}d\rv_{12}+
 \int n^{\xi}(\rv)\frac{\partial v^{\xi}(\rv)}
{\partial \xi}d\rv,
\label{eq_HFAPDFT}
\eeq
so that
\beq
G[f;V_{ne}]=T_{\rm f}[f]+\int_0^{\xi_{\rm phys}} d\xi\int d\rv\,
n^{\xi}(\rv)\frac{\partial v^{\xi}(\rv)}
{\partial \xi},
\label{eq_adiaf}
\eeq
where $T_{\rm f}[f]$ is the kinetic energy of a system of 
$N$ free (zero external potential)
interacting spin-$\frac{1}{2}$ 
fermions  having the same $f(r_{12})$ of the 
physical 
system. In the case of a confined system (atoms, molecules) the 
effective interaction $w^{\xi=0}(r_{12})$ must have an attractive tail:
the hamiltonian corresponding to $\xi=0$ in Eq.~(\ref{eq_Hxi}) describes
a cluster of fermions whose center of mass is translationally invariant.
The functional $T_{\rm f}[f]$ is the internal kinetic energy of
this cluster. 

\begin{figure}
\begin{center}
\includegraphics[width=9cm]{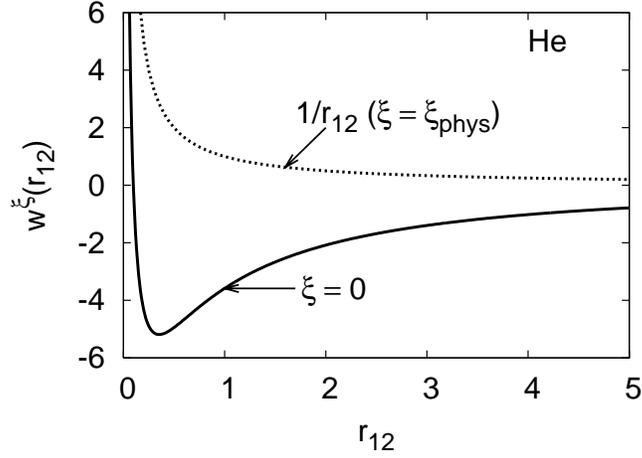} 
\end{center}
\caption{The electron-electron interaction at the two ends
of the APDFT adiabatic connection of 
Eqs.~(\ref{eq_Hxi})-(\ref{eq_fconst}) in the case of the He atom.}
\label{fig_weffHe}
\end{figure}

To fix the ideas, consider the simple case of two electrons, e.g. the He
atom. 
When $\xi=0$, we have two fermions in a relative bound state (similar
to the case of positronium, but with a different interaction). This 
relative bound state
is such that the square of the wavefunction for the relative coordinate
$r_{12}$ is equal to $f(r_{12})$ of the starting physical system.
The corresponding effective interaction 
$w^{\xi=0}(r_{12})$, obtained\cite{GS2} by inversion from a very accurate 
wavefunction,\cite{Umrig}
is shown in Fig.~\ref{fig_weffHe}, for
the case of the He atom.

For more than two electrons, $T_{\rm f}[f]$ is still a complicated
many-body object. Moreover, since 
the corresponding $w^{\xi=0}(r_{12})$ can have an attractive tail
(as in the case $N=2$),  we may have an ``exotic'' true ground-state 
for the cluster, i.e., the cluster state with the same $f(r_{12})$ of the
physical system can be an excited state. However, we have to keep in
mind that our aim is not to solve the many-electron problem by
means of APDFT alone: we want to use APDFT to produce realistic
$f(r_{12})$ along the DFT adiabatic connection 
of Eqs.~(\ref{eq_Hlambda})-(\ref{eq_nconst}). To this end, we 
proposed\cite{GS1,GS2} an approximation for the functional $T_{\rm f}[f]$
based on a geminal decomposition,
\beq
T_g[f]=\min_{\{\psi_i\}\to f}
\sum_i \vartheta_i\,\langle \psi_i|-\nabla^2_{r_{12}}|\psi_i\rangle,
\label{eq_Tg}
\eeq
where $\psi_i(r_{12})$ are some effective geminals (orbitals for
two electrons, but only depending on the relative distance
$r_{12}$), and $\vartheta_i$ some occupancy numbers to be chosen.
For example, one can always make a ``bosonic'' choice, by occupying
only one geminal,\cite{nagy2} equal to $\sqrt{f(r_{12})}$. The
geminals $\psi_i(r_{12})$ then satisfy the equations
\beq
\left\{\begin{array}{l}
[-\nabla^2_{r_{12}}+w_{\rm eff}(r_{12})] \psi_i(r_{12})  =  \epsilon_i\,
\psi_i(r_{12}) 
 \\ 
\sum_i \vartheta_i|\psi_i(r_{12})|^2  =  f(r_{12}),
\end{array}
\right.
\label{eq_eff}
\eeq
with
\beq
w_{\rm eff}(r_{12})=\frac{1}{r_{12}}+
\frac{\delta G[f]}{\delta f(r_{12})}-\frac{\delta T_g[f]}{\delta f(r_{12})}.
\eeq
The approximation of Eq.~(\ref{eq_Tg}) is mainly motivated by the need of
having simple equations for $f(r_{12})$ (the one-dimensional character of
Eqs.~(\ref{eq_eff}) is of course very appealing). Notice that only in the
case $N=2$ we have $T_g[f]=T_{\rm f}[f]$, and thus the effective
interaction $w_{\rm eff}(r_{12})$ of Eqs.~(\ref{eq_eff}) becomes equal to 
$w^{\xi=0}(r_{12})$.

\section{Combining DFT and APDFT in a self-consitent way}
As explained in the previous sections, to compute the expectation value
of the physical hamiltonian of Eq.~(\ref{eq_Hphys}) we only need
$f(r_{12})$ and $n(\rv)$ for $V_{ee}$ and $V_{ne}$, and 
the non-interacting KS kinetic energy $T_s[n]$ plus the APD
$f^\lambda(r_{12})$ along the DFT adiabatic connection (\ref{eq_Hlambda}) for
the expectation value of $T$; schematically:
\beq
\langle\Psi|H|\Psi\rangle=
\underbrace{\langle\Psi|T|\Psi\rangle}_{T_s[n]+f^\lambda(r_{12})}+\underbrace{\langle\Psi|V_{ee}|\Psi\rangle}_{ f(r_{12})}+\underbrace{\langle\Psi|V_{ne}|\Psi\rangle}_{n(\rv)}.
\eeq
Our aim is to obtain $n(\rv)$ and $T_s[n]$ via the KS equations,
and $f^\lambda(r_{12})$ via  Eqs.~(\ref{eq_eff}), which can be generalized
to any hamiltonian along the DFT adiabatic connection, 
by simply replacing the physical hamiltonian $H$ with $H^\lambda$ 
of Eqs.~(\ref{eq_Hlambda})-(\ref{eq_nconst}) in the steps
of Sec.~\ref{sec_APDFT}.

A self-consitent scheme for this construction reads
\begin{eqnarray}
& & \left(T+V_{\rm KS}\right)\Phi_{\rm KS}=E_{\rm KS} \Phi_{\rm KS}\qquad  
\qquad \qquad \Rightarrow \qquad
n(\rv),\;T_s[n] 
\label{eq_KSsche} \\
& & \left\{\begin{array}{l}
[-\nabla^2_{r_{12}}+w_{\rm eff}^\lambda(r_{12};[n])] \psi_i^\lambda(r_{12})  =  
\epsilon_i^\lambda\,\psi_i^\lambda(r_{12}) 
 \\ 
\sum_i \vartheta_i|\psi_i^\lambda(r_{12})|^2  =  f^\lambda(r_{12}),
\end{array}
\right.
\qquad   \Rightarrow  \qquad f^\lambda(r_{12}) 
\label{eq_efflambda}\\
& &  E_0=\min_{v_{\rm KS}} 
\left\{ T_s[n]+
\int_0^{\lambda_{\rm phys}} d\lambda\int d\rv_{12}
f^{\lambda}(r_{12})\frac{\partial w^{\lambda}(r_{12})}{\partial \lambda}
+\int d\rv\, n(\rv) v_{ne}(\rv)\right\}.
\label{eq_minvKS} 
\end{eqnarray}
The computation starts with a trial $v_{\rm KS}(\rv)$ in the KS
equations, schematically represented by Eq.~(\ref{eq_KSsche}), where
$\Phi_{\rm KS}$ is the Slater determinant of KS orbitals. From the KS equations
we thus get a first approximation for the density $n(\rv)$ and the
non-interacting kinetic energy $T_s[n]$.
Provided that we have a prescription to build an approximate 
$w_{\rm eff}^\lambda$ for a given density $n(\rv)$ 
(see next Sec.~\ref{sec_weff}), we can obtain $f^\lambda(r_{12})$
along the DFT adiabatic connection from
Eqs.~(\ref{eq_efflambda}). In general, this step is not
expensive: Eqs.~(\ref{eq_efflambda}) are unidimensional, and
if the dependence of $w^\lambda(r_{12})$ on $\lambda$ is smooth,
few $\lambda$ values (5-20) are enough to provide a good estimate
of the coupling-constant average. The physical ground-state energy
$E_0$ is then evaluated via Eq.~(\ref{eq_minvKS}). The procedure
should then be repeated by optimizing the KS potential
$v_{\rm KS}$, so that $E_0$ is minimum. The $N$-representability
problem of the KS exchange-correlation functional is clearly
shifted to the $N$-representability problem for $f^\lambda(r_{12})$.
In view of the new conditions derived for the pair density,\cite{ayersNPD} 
this seems to leave space for improvements.

\section{Properties of the effective electron-electron interaction}
\label{sec_weff}
So far, we have only replaced the problem of finding an approximation
for $E_{xc}[n]$ with the problem of constructing 
$w_{\rm eff}^\lambda(r_{12};[n])$.
In order to proceed further, we study here the properties of 
$w_{\rm eff}^\lambda(r_{12};[n])$. 

If we want  our Eqs.~(\ref{eq_KSsche})-(\ref{eq_minvKS}) to be fully
self-consitent, we should impose that for $\lambda=0$
 Eqs.~(\ref{eq_efflambda}) yield $f^{\lambda=0}(r_{12})=f_{\rm KS}(r_{12})$,
i.e., the same APD we would obtain by inserting the KS Slater determinant
of Eq.~(\ref{eq_KSsche}) in Eqs.~(\ref{eq_defP2})-(\ref{eq_deff}). This
corresponds, in the usual DFT language, to treat exchange exactly. The
first property we should thus impose to $w_{\rm eff}^\lambda(r_{12})$ is
\beq
w_{\rm eff}^{\lambda=0}(r_{12})=w_{\rm eff}^{\rm KS}(r_{12}).
\label{eq_wKS}
\eeq
If we use only one geminal to define $T_g[f]$ in Eq.~(\ref{eq_Tg}),
the property (\ref{eq_wKS}) corresponds to $w_{\rm eff}^{\lambda=0}(r_{12})=\nabla^2 \sqrt{f_{\rm KS}(r_{12})}/
\sqrt{f_{\rm KS}(r_{12})}$. For more than one geminal we need
more sophisticated constructions, mathematically equivalent
to those used to construct the KS potential $v_{\rm KS}(r)$ for
a given spherical density $n(r)$.\cite{colonna}
Equation~(\ref{eq_wKS}) also provides a very good starting
point to build $w_{\rm eff}^\lambda(r_{12})$: the KS system already
takes into account the fermionic structure and part of the effect
of the external
potential in the physical problem. What is then left, that needs to be 
approximated, is the effect of turning on the electron-electron interaction
without changing the one-electron density $n(\rv)$, and the difference
between $T_{\rm f}[f]$ and $T_g[f]$.

For confined systems (atoms, molecules) another property to be imposed
on $w_{\rm eff}^\lambda(r_{12})$ concerns the eigenvalue
$\epsilon_{\rm max}^\lambda$ corresponding to the highest occupied geminal
in Eqs.~(\ref{eq_efflambda}). 
In fact, the asymptotic behavior of the pair density 
$P_2(\rv_1,\rv_2)$ of Eq.~(\ref{eq_defP2}) for $|\rv_1|\to\infty$
(or $|\rv_2|\to\infty$) is, in this case,\cite{P2longrange}
\beq
\lim_{|\rv_1|\to\infty}P_2(\rv_1,\rv_2)=n(\rv_1)n_{N-1}(\rv_2)\{\hat{\rv}_1\},
\label{eq_P2larger}
\eeq
where $n_{N-1}(\rv)$ is one of the degenerate ground-state densities 
of the $(N-1)$-electron system (in the same external potential
$V_{ne}$), with the choice depending parametrically
upon the direction $\hat{\rv}_1=\rv_1/|\rv_1|$. A similar expansion
holds for the KS pair density, obtained from the KS Slater determinant
of Eq.~(\ref{eq_KSsche}),
\beq
\lim_{|\rv_1|\to\infty}P_2^{\rm KS}(\rv_1,\rv_2)=n(\rv_1)n_{N-1}^{\rm KS}(\rv_2)\{\hat{\rv}_1\},
\label{eq_P2KSlarger}
\eeq
where we have used the fact that, by construction, the $N$-electron density
is the same for the physical system and for the KS one (while, of course,
the corresponding $(N-1)$-electron densities are in general different).
For a given attractive (atomic, molecular) external potential vanishing
at large distances,
the $N$-electron density is in general more diffuse (decaying
slower at large distances) than the $(N-1)$-electron
density, so that the 
asymptotic behavior of the APD $f(r_{12})$ is, for large $r_{12}$, 
dominated by the $N$-electron density decay at large distances. 
We thus see, from Eqs.~(\ref{eq_P2larger}) and
(\ref{eq_P2KSlarger}), that the corresponding APD's, $f(r_{12})$ and
$f_{\rm KS}(r_{12})$, will have the same
large-$r_{12}$ decay, $\propto e^{\sqrt{-2\epsilon_{\rm max}}\,r_{12}}$, with
a different prefactor (which can also include a polynomial function of
$r_{12}$), depending on the difference between $n_{N-1}(\rv)$
and $n_{N-1}^{\rm KS}(\rv)$. Since the same expansion holds for any
$P_2^\lambda(\rv_1,\rv_2)$ along the DFT adiabatic connection,
\beq
\lim_{|\rv_1|\to\infty}P_2^\lambda(\rv_1,\rv_2)=n(\rv_1)n_{N-1}^\lambda(\rv_2)\{\hat{\rv}_1\},
\label{eq_P2lambdalarger}
\eeq
the highest eigenvalue $\epsilon_{\rm max}^\lambda$ in 
Eqs.~(\ref{eq_efflambda})
must be independent of $\lambda$ and equal to the one for the
KS APD, 
\beq
\epsilon_{\rm max}^\lambda=\epsilon_{\rm max}^{\lambda=0}=\epsilon_{\rm max}^{\rm KS}.
\eeq
In particular, if we choose only one geminal\cite{nagy2} for the definition of
$T_g[f]$, there is only one eigenvalue, which must be the same for every
$\lambda$.

For an extended system
we have scattering states in Eqs.~(\ref{eq_efflambda}). For the 
special case of the uniform electron gas, Eqs.~(\ref{eq_efflambda})
become equivalent to an approach that was first proposed by
Overhauser,\cite{Ov} and further developed by other authors 
in the past five years.\cite{GP1,DPAT1,CGPenagy2} In this approach, the
geminal occupancy numbers $\vartheta_i$ are the same as the ones for
a Slater determinant: occupancy 1 for singlet states (even relative
angular momentum $\ell$), and occupancy 3 for triplet states
(odd relative angular momentum $\ell$), up to $N(N-1)/2$ geminals.
Rather simple approximations for the 
effective potential $w_{\rm eff}^\lambda(r_{12})$ gave good 
results\cite{GP1,DPAT1,CGPenagy2} for the radial distribution
function $g(r_{12})$, when compared with quantum Monte Carlo data. 
The long-range asymptotic behavior, in this case, 
corresponds to a phase-shift sum rule for the interaction 
$w_{\rm eff}^\lambda(r_{12})$.\cite{ziescheFriedel} 
The choice of one geminal for the uniform electron gas
has been explored, with remarkable success, in Ref.~\refcite{pisani1gem}.
In this case, the formal similarity with the Fermi-hypernetted-chain
approach\cite{krot1} (FHCN) was exploited to build a good approximation for
$w_{\rm eff}^\lambda(r_{12})$ (which was  split into $\uu$
and $\ud$ contributions).

Finally, the small-$r_{12}$ behavior of the effective potential 
$w_{\rm eff}^\lambda(r_{12})$  is determined by the choice of the
adiabatic connection path. For instance, if we choose $w^\lambda(r_{12})
=\lambda/r_{12}$ in Eqs.~(\ref{eq_Hlambda})-(\ref{eq_defTs}), then
the APD $f^\lambda(r_{12})$ displays the electron-electron cusp
$f^\lambda(r_{12}\to 0)=f^\lambda(0)(1+\lambda\, r_{12}+...)$, which
implies, in turn, that also $w_{\rm eff}^{\lambda}$ must behave,
{\it for small $r_{12}$}, as $w_{\rm eff}^{\lambda}(r_{12}\to 0)
= \lambda/r_{12}+...$.
In particular, for $\lambda=\lambda_{\rm phys}$, we always have
$w_{\rm eff}^{\lambda_{\rm phys}}(r_{12}\to 0)
= 1/r_{12}+...$, as shown, for the case of the He atom
in Fig.~\ref{fig_weffHe}. 
If we choose a cuspless nonlinear path, like
$w^\lambda(r_{12})=\erf(\lambda\,r_{12})/r_{12}$, then the
small $r_{12}$ behavior of $w_{\rm eff}^\lambda(r_{12})$
is known only when we are approaching the physical interaction.\cite{GS3}

\section{Preliminary applications}
The construction of an approximate $w_{\rm eff}^\lambda(r_{12};[n])$ can 
thus start with the decomposition
\beq
w_{\rm eff}^\lambda(r_{12};[n])=w_{\rm eff}^{\rm KS}(r_{12})+
w^\lambda(r_{12})+\Delta w_{\rm eff}^\lambda(r_{12};[n]),
\eeq
where the term $\Delta w_{\rm eff}^\lambda(r_{12};[n])$
should take care of the fact that, when the electron-electron
interaction is turned on, the one-electron density $n(\rv)$
and (for confined systems) the highest
eigenvalue $\epsilon_{\rm max}^\lambda$ do not change.

As a starting point, we applied the method 
of Eqs.~(\ref{eq_KSsche})-(\ref{eq_minvKS}) to the He isoelectronic series.
In this simple (yet not trivial) 2-electron 
case, we have the advantage that we can
treat $T_{\rm f}[f]$ exactly. We developed an approximation\cite{GS1} for 
$\Delta w_{\rm eff}^\lambda(r_{12};[n])$ based on the one used for
the uniform electron gas.\cite{GP1} This approximation is designed
to mimic the conservation of $n(\rv)$ along the DFT adiabatic connection,
but does not take into account the eigenvalue conservation. It
works remarkably well when combined with a nonlinear adiabatic
connection path
$w^\lambda(r_{12})=\erf(\lambda\,r_{12})/r_{12}$ that separates
short- and long-range correlation effects, and is reported in the Appendix
of Ref.~\refcite{GS1}. Preliminary implementations of the self-consitent
procedure of Eqs.~(\ref{eq_KSsche})-(\ref{eq_minvKS}) yield
ground-state energies within 1~mH with respect to
full configuration interaction (CI) calculations in the same basis set.
However, the way we carried out these first tests was simply based on a
direct minimization of few variables parametrizing $v_{\rm KS}(\rv)$.
This rather inefficient way to implement 
Eqs.~(\ref{eq_KSsche})-(\ref{eq_minvKS})
needs further improvment. Besides,
the Kohn-Sham potentials we obtain in this way are very unstable
(like the ones of Ref.~\refcite{scuseria}), although the 
corresponding total energies and
electronic densities are stable, and do not display the variational
collapse of perturbation-theory-based approximate $E_c[n]$.
\cite{variationalcollapse} 

\section{Perspectives}
The generalization to many-electron systems of nonuniform density 
of the approximation built in Ref.~\refcite{GS1} 
for $w_{\rm eff}^\lambda(r_{12})$ is not straightforward. If we simply apply
it to the Be atom case (by using only one geminal), we obtain energy
errors of 300~mH. Adding the eigenvalue conservation can improve the 
results, but, of course, there is not a unique way to impose it. So far,
we found that the final outcome strongly depends on how we impose the 
eigenvalue conservation (i.e., if we use a functional form with one
parameter adjusted to keep the eigenvalue independent of
$\lambda$, the results drastically depend on the chosen functional form).
It seems thus necessary to switch to more than one geminal, and/or
to find better constructions for $w_{\rm eff}^\lambda(r_{12})$.

In particular, it may be promising to explore the possibility to
construct approximations inspired to the FHNC,\cite{pisani1gem,krot1,krot2}
and to try to include some of the new results on $N$-representability
conditions for the pair density.\cite{ayersNPD} Different
approximations with respect to the one of Eq.~(\ref{eq_Tg}) for the functional
$T_{\rm f}[f]$, and the use of Eq.~(\ref{eq_adiaf}) also deserve further 
investigation.
 
\section*{Acknowledgments}
We thank E.K.U. Gross for useful discussions and suggestions. One
of the authors (P.G.G.) gratefully aknowledges the {\em 30th 
International Worksohop on Condensed Matter Theory} organizers for
supporting her participiation to the meeting.



\end{document}